# Anionic hydrogen clusters as a source of diffuse interstellar bands (DIBs)[1]

Lulu Huang,[1] Chérif F. Matta*,[2-5] Daniel Majaess,[2] Tina Harriott,[2,6] Joseph Capitani,[7] and Lou Massa*[1]

[1] *Hunter College and the Graduate School, City University of New York, New York, NY 10065, USA.* [2] *Department of Chemistry and Physics, Mount Saint Vincent University, Halifax, Nova Scotia, Canada B3M2J6.* [3] *Department of Chemistry, Saint Mary's University, Halifax, Nova Scotia, Canada B3H3C3.* [4] *Dép. de chimie, Université Laval, Québec, Québec, Canada G1V 0A6.* [5] *Department of Chemistry, Dalhousie University, Halifax, Nova Scotia, Canada B3H4J3.* [6] *Department of Mathematics and Statistics, Mount Saint Vincent University, Halifax, Nova Scotia, Canada B3M2J6.* [7] *Chemistry and Biochemistry Department, Manhattan College, Riverdale, NY 10471, USA.*
\* Correspondence: cherif.matta@msvu.ca, lmassa@hunter.cuny.edu

**TOC GRAPHIC**

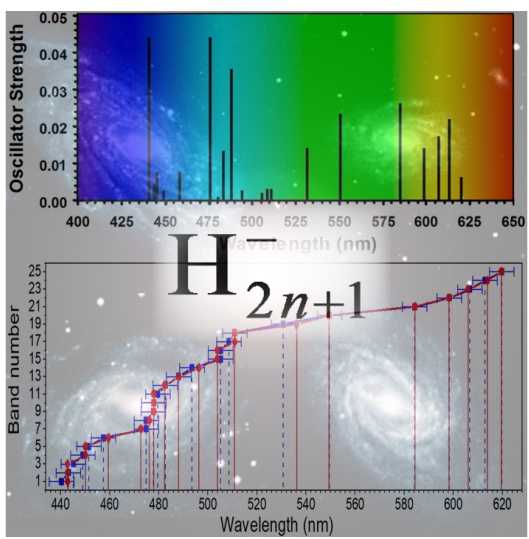

**ABSTRACT**
The sources behind numerous key diffuse interstellar bands (DIBs) remain elusive, and thus evidence is presented here that a family of seven anionic hydrogen atomic clusters ($H_{2n+1}^-$, 1≤$n$≤7) are pertinent contributors. Configuration interaction calculations sho hat these charge-induced dipole clusters are stable at temperatures characteristic of the interstellar medium, and emerge from the most abundant element in the Universe. The clusters' spectra yield 25 absorption optical lines that align with observed DIBs to within the computational uncertainties. The absorption bands are due to excitations from the ground states of the clusters to metastable states.

## 1. INTRODUCTION

Light from stars and galaxies may be absorbed by interstellar material along the sight-line. Absorption typically affects the entire spectrum rather than producing lines. However, for over 100 years a set of discrete absorption features, known as diffuse interstellar bands (DIBs), have been recognized in the spectra of many stars when viewed through interstellar clouds. Indeed, the signature of that effect is partly conveyed by numerous discrete absorption features known as diffuse interstellar bands (DIBs). Yet there is no consensus as to the source(s) behind the majority of DIBs, and hence the impetus behind this research.

For a comprehensive discussion concerning the discovery of DIBs the reader is referred to McCall & Griffin (2013) [1]. Briefly, Hartman 1904 [2] noted that the K line of ionized calcium did not exhibit the requisite Doppler shift identified in other lines associated with the spectroscopic binary δ Orionis, and consequently it was associated with an interstellar calcium cloud along the sightline. In 1919 Heger reported that sodium D lines displayed a similar stationary behaviour relative to other lines, and later identified features at 5780 Å and 5797 Å that are now recognized as DIBs [1]. That work in part helped catalyze this key area of research, and significant progress has since been achieved toward expanding the census of DIBs. In 2009, Hobbs *et al.* [3] reported a total of 414 DIBs between 3900 Å and 8100 Å in the spectrum of the Galactic binary star HD 204827. DIBs were likewise detected beyond the Galaxy, such as at 5780 Å and 5797 Å in the Magellanic clouds [4], and Junkkarinen *et al.* [5] found DIBs associated with observations of the BL Lac object AO 0235+164 ($z = 0.524$).

The word "diffuse" in the designation of DIBs underscores the "somewhat hazy appearance of DIBs compared with the relative sharpness of atomic transitions in the interstellar medium", quoting McCall & Griffin (2013) [1], which they argue is a strong indicator that the sources of DIBs are molecular (not atomic). The diffuseness of the DIBs is generally attributed to the short lifetimes of the excited molecular species that are associated with them (See Ref. [6] and references therein). The observed widths of DIBs generally fall between 0.05 and 0.3 nm, with a typical width of ~ 0.07 nm according to a review by Tielens [7].

Research continues unabated to identify the numerous sources behind DIBs, as the strengths of DIBs are not necessarily correlated. DIB wavelengths are unrelated to known atomic transitions, and may be linked to molecules or molecular clusters. The existence of spectroscopic fine-structure in DIBs supports that conclusion, namely that the source is molecular, since the fine-structure can result from molecules composed of different isotopes of certain atoms such as carbon 12 and carbon 13. The ESO Diffuse Interstellar Bands Large Exploration Survey (EDIBLES) highlights in excess of 500 DIBs [8], however $C_{60}^+$ remains the sole confirmed source for a few DIBs. For example, Maier *et al.* [9] advocated that fullerene $C_{60}^+$ was the source of DIBs at 9577 Å and 9632 Å, and three other $C_{60}^+$ bands were since identified at 9348 Å, 9365 Å, and 9428 Å [10].

A source for certain DIBs that features hydrogen is favored by the element's abundance. Yet atomic (H) or molecular ($H_2$) hydrogen are ruled out as sources of DIBs because their spectra do not match known DIBs. Moreover, $H_2$ is rare in the interstellar medium [11]. The H- is also not a viable candidate since it has only one bound state, at least within the non-relativistic, fixed nucleus approximation with Coulombic forces only (and no excited bound states) [12].

As a result, it has been suggested that ion induced dipole clusters $H_{2n+1}^-$ (where *n* is the number of $H_2$ ligands) might exist in the interstellar medium [13-17], and could be a viable source of DIBs. While the H$^-$ ion is found in the interstellar medium CHEMREV and it is also found in stellar atmospheres and regions such as dark molecular clouds [18]. Indeed the ions $C_4H^-$, $C_6H^-$, $C_8H^-$ are among the reported 2018 census [19] of 204 molecules identified in the interstellar medium. Even though the binding energy of H$^-$ is low (~ 0.75 eV), the low temperatures (~ 7-20 K) and densities (~ 100-300 molecules/cm$^3$) of certain molecular clouds imply that once formed H$^-$ could remain stable for long periods, and act as an attractor for $H_2$ molecules to form anionic hydrogen atomic clusters.

This work investigates anionic hydrogen clusters as contributors to the DIBs. Anions, in general, are believed to be contributing to the DIBs spectrum [20]. Moreover, Sarre has argued that anion-dipole bound molecular clusters, such as those presented here, are good candidates as a source of DIBs [21]. This paper is divided as follows, in Section 2 the computational approach is outlined. In Section 3.1 the evidence for the stability of these clusters under interstellar conditions is presented along with a mechanism for their formation. Finally, the spectra of these clusters and the matching DIBs are discussed in Section 3.2.

## 2. CALCULATIONS

Configuration interaction calculations with single excitations (CIS) were conducted for the ground and excited states. Brillouin's theorem states that the matrix element of a singly excited Slater determinants (SD) with the ground state SD vanishes ($\left\langle ^0\Psi_{SD} \middle| \hat{H} \middle| ^1\Psi_{SD} \right\rangle = 0$) [22]. The singly excited states, thus, cannot improve the wavefunction or energy of the ground state. Thus, in this work the CIS calculations are followed by Møller-Plesset second-order perturbation theory (MP2) corrections [23,24] (with frozen core) for the ground and excited states, a method referred to as CIS-MP2.

The triple-zeta quality aug-cc-pVTZ basis set has been used in all calculations whereby the geometry of any given cluster is energy minimized without constraints at the CIS/aug-cc-pVTZ level of theory followed by a harmonic frequency calculation. No imaginary frequencies were found for any of the optimized geometries, and hence all structures are (at least) local minima on their respective potential energy hypersurfaces. The wavelengths of maximum absorption ($\lambda_{max}$) were calculated at the CIS-MP2/aug-cc-pVTZ//CIS/aug-cc-pVTZ level of theory. This theoretical methodology has been found to reproduce UV-Vis spectra accurately (even with smaller basis sets) by Forseman, Head-Gordon, and Pople [23]. In the present work, the basis set aug-cc-pVTZ was selected since it is relatively large, correlation consistent, and – importantly – it is augmented with diffuse functions that are necessary for an accurate description of anionic systems, especially in their excited electronic states.

The calculated spectral lines correspond to $\Delta E = h\nu$, the difference between the MP2 energy of the ground state and the energy of the excited state after the MP2 correction is added to account for electron correlation. Binding energies are defined as usual:

$$E_{binding}(n) = E_{cluster}(n) - E_{H^-} - nE_{H_2}, \quad (1)$$

where all symbols have transparent meanings.

Atomic charges were obtained in accordance with the quantum theory of atoms in molecules (QTAIM) [25-27] using the AIMAll/AIMStudio software [28].

## 3. RESULTS AND DISCUSSION

### 3.1 Structural Stabilities and Lifetimes of the Anionic Hydrogen Clusters

*3.1.1. Binding and Electron Detachment Energies, Mean Lifetimes, and Mechanism of Formation in Interstellar Space:* Ball-and-stick representations of the optimized geometries of the clusters are displayed in Fig. 1 and their properties are listed in Table 1. From the table one observes a narrowing of the HOMO-LUMO gap with the size of the cluster up to and including $H_{11}^-$ after which it reaches an asymptote in the vicinity of – 0.1620 atomic units (a.u.) ($\approx$ 4.41 eV). The magnitudes of the binding energies increase with addition of H$_2$ ligands, as previously observed at the DFT level of theory [29], but with $H_{11}^-$ falling outside of the trend (this outlier's geometry also differs from other clusters in that not all of its H$_2$ ligands radiate from the central anion – See Fig. 2).

The increase in the binding energy of these clusters with size suggests that the search for larger hydrogen ion clusters ($n > 7$) is indicated theoretically and experimentally. Cationic analogs of these clusters, $H_n^+$, are known to exist for all values of $3 < n$ (odd) $< 99$ [30], hence it is not unreasonable to expect the existence of anionic clusters as well [31,32]. Recently, path integral molecular dynamics simulations at 1 K by Calvo and Yurtsever have pushed the search for stable anionic clusters up to $n = 54$ molecular hydrogen ligands attached to a central hydride [33]. These workers found that stable icosahedral shells of the larger ions exhibit magic numbers at sizes of 12, 32, and 44 ligands (confirming experimental results previously published by Renzler *et al.* [34]) [33].

CCSD//MP2 calculations by Renzler *et al.* with quadruple zeta quality basis set of the (vertical) electron affinities of the neutral species (H, H$_3$, H$_5$, and H$_7$) yield 0.74, 0.77, 0.81, and 0.69 eV, respectively. Given that the excitation energies to the first excited state of all clusters in this work are between 4 and 5 eV, *i.e.* larger in magnitude than the electron detachment energies, the excited states are metastable and short-lived auto-ionization (Auger) states which introduces line broadening through Heisenberg's uncertainty principle.

Binding energies range from -0.02 eV (-0.4 kcal/mol) for the $H_9^-$ cluster to -0.22 eV (-5.2 kcal/mol) for the $H_{15}^-$ cluster. The binding energies are more negative (more stabilizing) with the number of ligands at an average rate of *ca.* -0.03 eV per ligand (excluding $H_9^-$) from a mere -0.05 eV ($H_3^-$) to -0.22 eV ($H_{15}^-$). The excitation energies being higher than these values, that can also imply that the calculated excited states are metastable in this respect as well being auto-dissociating states, which can shorten their life-time further and introduces more broadening to the lines. This broadening is at the moment out of the scope of this preliminary study and will be the subject of a follow up study in the near future.

Most hydrogen in dense dust clouds is molecular since carbon-, silicon-, and/or mineral-based dust particles (M) operate as condensation centers increasing the effective concentration of the reactants and providing a collision recoil energy absorbing support [11]. Accordingly, the formation of molecular hydrogen can be represented by [11]:

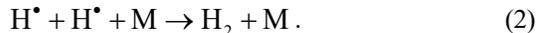

$$H^\bullet + H^\bullet + M \rightarrow H_2 + M. \quad (2)$$

It is hereby suggested that a similar mechanism might lead to the formation of the anionic clusters examined in the present work.

This would involve the capture of a "*rogue electron* [35]" by atomic hydrogen (the electron affinity of which is 0.75 eV, compared to the thermal energy $E_{\text{therm.}} \equiv k_B T \approx 0.002$ eV at 20 K) and the physisorption of the resulting hydride on a dust particle M. A sequential addition of molecular hydrogen formed nearby according to Eq. (2) will result into the proposed clusters (Eq. (3) below). Symbolically, we may write:

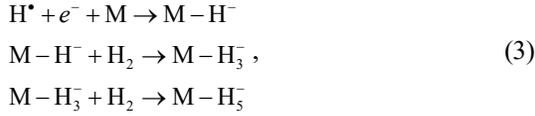

$$\begin{aligned} &H^\bullet + e^- + M \rightarrow M - H^- \\ &M - H^- + H_2 \rightarrow M - H_3^-, \\ &M - H_3^- + H_2 \rightarrow M - H_5^- \end{aligned} \qquad (3)$$

etc.

The thermal energy determines the stability of the complexes and their average lifetime. In the Milky Way's galactic plane, dust of size of the order of 140-240 μm has temperatures of about 16 to 21 K depending on location within the galaxy according to the Cosmic Background Explorer (COBE) satellite data (See Table 7.12, p. 160 of Ref. [36] ). Under these conditions (~ 20 K) the thermal energy is *ca.* 0.002 eV (0.04 kcal/mol), an order of magnitude lower than the binding energy of the least bound cluster ($H_9^-$, -0.02 eV). To provide other reference thermal energies, for a hotter cloud at, say, 100 K, this energy is ~ 0.009 eV (0.20 kcal/mol) (which is still lower than the binding energy of the least bound cluster) while at room temperature (298 K) it is around 0.026 eV (0.59 kcal/mol).

With binding energies, the ambient thermal energy, and vibrational frequencies, and using the Wigner-Polanyi equation, we can estimate the characteristic lifetime expectancy of these complexes ($\tau$) *in the ground state*, whereby:

$$\tau = \frac{1}{\nu} \exp\left(\frac{|E_{\text{bind}}|}{RT}\right), \qquad (4)$$

in which the magnitude of the binding energy is the minimum energy needed to dissociate the cluster (Table 1).

The interpretation of the pre-exponential factor in the Wigner-Polanyi equation - that has the dimensions of frequency - is not entirely clear theoretically. This frequency factor is usually an experimentally determined parameter obtained from programmed thermal desorption. We may, however, adopt the interpretation of Chaabouni *et al.* [37] who interpret this factor as the "*attempt frequency of molecules to overcome the barrier*" (of dissociation). The highest vibrational frequency is taken as this "*attempt frequency*" since this will limit the lifetime to its shortest possible value (since it represents the fastest movement that can bring about a dissociation event).

Table 1 lists the fastest calculated (harmonic) vibrational frequencies of the seven clusters and their corresponding characteristic lifetimes. From the listed lifetimes, and since electronic transitions occur at the femtosecond time-scale [38], it is clear that even at room temperature all seven clusters can be sufficiently stable to absorb/emit in the UV-Vis. (Even the shortest lived species, $H_9^-$, would still survive 10 fs, while $H_{15}^-$ - exhibiting the longest lifetime $\tau$- is predicted to last $10^5$ fs).

More importantly, at the temperature of the interstellar clouds (taken as 20 K), five of the clusters are stable with practically infinite lifetimes, while the remaining two, namely $H_3^-$ and $H_9^-$, have $\tau \sim 0.05$ s and $\sim 10^5$ fs, respectively, which is sufficiently long to absorb/emit in the electronic transition region. Thus, all seven clusters are predicted to survive for times many orders of magnitudes longer than the time-scale of electronic transitions and are potential contributors to the spectrum. These lifetimes would be exponentially longer at the interstellar space background temperature of 3 K. The calculated lifetimes are consistent with the low temperature experimental determinations of the lifetimes of $H_n^-$ ($n = 3$) by Wang *et al.* [39] and ($n \geq 5$) by Renzler *et al.* [34] *et al.* which are at least 8 μs – the time necessary to fly through the mass spectrometer.

The excitation of the clusters to their above-mentioned auto-ionization/auto-dissociation states is accompanied by broadened absorption lines centered at the predicted frequencies/wavelengths. The metastable excited states can eject an electron with a kinetic energy equal to the gap between the excited state of the anion and the ground state of the neutral species. Another possible channel is that the metastable state dissociates into a hydride ion and neutral hydrogen molecules sharing the energy difference (between the excited state of the anion and the sum of the energies of the components at infinite separation) as their translational, rotational, and vibrational energies.

It is, hence, predicted that absorption by one of these clusters is a one-time event after which the cluster falls apart through either auto-ionization or auto-dissociation (with line broadening depending on the lifetime of the metastable state). Essentially, the DIBs that are proposed to originate from these clusters are the result of a dynamic process. This process involves the constant formation of these clusters followed by a single event of excitation/absorption that leads to auto-ionization or auto-dissociation. In other words, the clusters are continually forming and breaking and, in the process, give rise to the observed bands.

*3.1.2 Geometries and Charge Distributions:* The energy-minimized geometries are perhaps not intuitively obvious candidate equilibrium structures, as they do not always adopt the highest possible symmetry. They all do however correspond to true energy minimum geometries. A search over the molecular energy hypersurface did find, in every case, a geometry which satisfies the criterion for equilibrium, *viz*., all vibrational frequencies are real (no imaginary frequencies found).

These optimized geometries for the $H_{2n+1}^-$ species (Fig. 1) seem to violate VSEPR (Valence Shell Electron Pair Repulsion Model) [40] rules if the cluster species are described using the standard Nyholm-Gillespie $AB_nE_m$ classification scheme, with E representing lone pairs of electrons. (In this case, $m = 1$.) However, owing to the "central atom" being the hydride H$^-$ ion, and the weak attraction of the ligand $H_{2n}$ species, VSEPR-like rules can be recovered by denoting the central hydride as AE, and the remaining $H_{2n}$ ligands as unified pseudo-atoms $B_n$, so that the clusters are now considered as $(AE)B_n$ species. In this way, each cluster geometry now shows general agreement with VSEPR-like rules. Any "fish tailing" in the terminal ligand H could be due to the longer ligand – central hydride bond lengths and merits additional study.

The atomic charges derived from QTAIM [25-27] are given in Table 2. Atomic charges alternate between negative and positive values (H$^-$...H$^{\delta+}$–H$^{\delta-}$) and the central anion (H$^-$) is found to carry most of the negative charge (from – 0.96 $e^-$ in $H_3^-$ to – 0.91 $e^-$ in $H_{15}^-$). The remainder of the negative charge is carried by the hydrogen atom(s) farthest from the central anion. The $H_2$ ligands are all polarized with an internal charge transfer around 0.1 $e^-$ with the positively-charged hydrogen atom being that nearest to the central anion. That hydrogen atom and the central anion share a bond of intermediate strength termed the "tri-hydrogen bond" [29], a variant of the dihydrogen bond [41] in which both the proton donor and acceptor are hydrogen atoms. The principal chemical bonding interaction holding these clusters together is, hence, of the ion–induced dipole type.

*3.1.3. Anionic Hydride-Centered Clusters and their Similarity to Corresponding Halide Centered Clusters:* Density functional theory (DFT) [42-44] B3LYP/6-311++G(*d*,*p*) calculations led to the identification of the equilibrium geometries of these clusters with the formula $H_{2n+1}^-$ where *n* is the number of $H_2$ ligands as well [29,45]. The clusters consists of a central hydrogen anion (H$^-$) and a number 1 ≤ *n* ≤ 6 of polarized $H_2$ ligand(s) "radiating" from the central ion like a pin cushion [29,45].

A set of anionic polybromide clusters, of similar stoichiometries and geometries to the hydrogen clusters reported here, have been reported by Feldmann [46] and theoretically investigated by Pichierri [47]. There is a X$^-$(H$_2$)$_n$ [48] and Br$^-$(Br$_2$)$_n$ (X=F, Cl, Br) [47] that corresponds to every one of the seven anionic hydrogen clusters H$^-$(H$_2$)$_n$ subject to this study[29]. Pichierri reports symmetries, geometries, and atomic and bond properties of a set of $Br_{2n+1}^-$ (1 ≤ *n* ≤ 6) clusters (Br$_2$ ligands bound to a central Br$^-$ ion), all of which are similar to the anionic hydrogen clusters reported here and elsewhere [29,45]. The one-to-one correspondence between the anionic hydrogen cluster and the corresponding anionic bromine clusters may suggest that *hydrogen with its a single valence-shell electron behaves, in this context, similarly to bromine with its single valance-shell hole.*

## 3.2 The Anionic Cluster's Coincident Spectra with Observed Diffuse Interstellar Bands (DIBs)

Anionic hydrogen clusters [14-17] are stable at the low temperatures and pressures endemic to the interstellar medium, and they are formed from the most abundant element in the Universe. There exists a plausible mechanism for their formation through condensation over interstellar dust particles, and we now show that their calculated spectra are consistent with observed lines in the DIBs.

Inasmuch as these anionic clusters have energy minima, lines of their visible spectra were calculated and compared to known nearby diffuse interstellar bands (DIBs) [3,49-54]. The calculated absorption spectra of this set of seven clusters exhibit, cumulatively, 25 lines align with the observational DIBs to within the uncertainties. The lines, reported below, are of particular interest because negative hydrogen ions $H^-$ and molecular hydrogen $H_2$, the reactants that produce the proposed clusters, are available within the interstellar medium [55].

The calculated maxima in the visible spectra of these clusters are listed along with their intensities and the nearest observed lines in the DIBs spectrum in Table 3. The position and intensity of the calculated spectral lines against a background of corresponding visible colors is shown in Fig 2. The calculated and experimental $\lambda_{max}$ values are plotted against a DIBs' number-designation of increasing wavelengths in Fig. 3. The figure also displays uncertainties estimated for each calculated DIB. The formula connecting the calculated wavelength to its corresponding excitation energy is:

$$\lambda \text{ [nm]} = 1{,}241 / E \text{ [eV]}. \qquad (5)$$

The uncertainties calculated from this relation are:

$$d\lambda \text{ [nm]} = -1{,}241 \, dE \text{ [eV]} / E^2 \text{ [eV}^2\text{]}. \qquad (6)$$

The qualitative estimate of the calculated uncertainty magnitudes are taken from the work of Forseman, Head-Gordon, and Pople (F.H.-G.P.) where the uncertainties associated with the CIS+MPE chemical model employed here are analyzed [23]. For the test molecule ethylene, and using a 6-31+G*R basis set (the R in this stands for "Rydberg" whereby non-standard extremely diffuse basis functions were added for the proper description of Rydberg states), F.H.-G.P. report uncertainties for a variety of transitions falling in the range: 0.01 eV < $dE$ < 0.05 eV [23]. These authors also found that, as the basis set is enlarged, there corresponds a consistent reduction in the calculated excitation energy uncertainties. As the basis used in the present work (aug-cc-pVTZ) is larger than that of Ref. [23] and is correlation consistent, a conservative assumption would be that our calculated uncertainties fall midrange of the above-mentioned uncertainty range. That is to say, a reasonable estimate of the uncertainty in our calculations of the wavelengths of maximum absorption is $dE \approx 0.025$ eV.

The calculated excitation energies of the hydrogen clusters examined in the present study all fall in the range $2.22 < E < 2.75$ eV. Thus, to calculate the qualitative uncertainties according Eq. (6), we take a midrange value of $E = 2.5$ eV. The resulting bars of calculation uncertainty have magnitude $d\lambda$ [nm] = ± 5 nm. The observational uncertainties are listed in Table 3 for each observation and are generally more than an order of magnitude less than for the calculated wavelengths.

Seven of the wavelengths listed in Table 3 are considered to be *definitive matches* between calculation and observation and are indicated in Table 3 by asterisks. In each of these cases, the observed line, is the *only* observed line that falls within the error bounds surrounding the calculated wavelength. The remaining observed wavelengths all fall within the error bars of the corresponding calculated wavelengths, but not uniquely so. That is to say, for those cases (without the asterisks in Table 3) there exist more than one observed wavelength within the calculated error bars. So, in those cases, the match of a calculated wavelength with a single observed wavelength partner is not unique. The listed observed wavelengths in the table, other than the seven definitive matches, are those which are closest to the corresponding calculated value. The computational uncertainty bars are displayed in Fig. 3. Both Table 3 and Fig. 3 suggest that, significantly, these clusters collectively exhibit optical spectral lines which are coincident with observed DIBs within the uncertainties. It is likewise notable that three of the clusters studied ($n'$ = 9, 13, and 15) have lines close to the strong experimental DIBs of $\lambda$ = 442.89 nm [50].

## 4. CONCLUSIONS

A critical outstanding problem in astronomy is the chemical source of DIBs, and in this work anionic hydrogen clusters are put forth as viable contributors, namely on the basis of several lines of evidence. Anionic hydrogen clusters, held together by anion induced dipolar bonds, stem from material available in interstellar space. Quantum mechanical computations suggest that these clusters are stable as they exhibit no imaginary frequencies, and possess a net binding energy sufficient for holding them together at temperatures characteristic of interstellar clouds (e.g., 20 K). The clusters are stabilized by an alternation of atomic charges ($H^-...H^{\delta-}-H^{\delta+}$) consistent with previous calculations by Huang *et al.* [29,45], and with corresponding bromine clusters [47]. It is also shown that the hydride ion and its $H_2$ ligand(s) adopt optimized geometries that obey VSEPR-like rules. These structures are similar to previously reported halogen-based clusters [47,48], whereby an electron pair in the hydrogen clusters appears to play a the role of a hole in determining the geometry of the bromine clusters [47].

Most importantly, the computed optical absorption spectra of negative anionic clusters match numerous observed DIBs, to within the uncertainties. While H⁻ itself has been proposed as a source of DIBs in the past, such claims were dismissed experimentally and computationally (see the discussion and references in Rau's review article [56]). What is proposed here are H⁻-centered clusters, that is, $H_2$-"solvated" hydride anions as a source of these bands. These anionic clusters are metastable in their electronic excited states. The energies of the lowest excited states lie above both the threshold of electron detachment and the dissociation of the clusters into their constituent fragments. Hence, the excited states of the clusters are short-lived. To estimate the broadening of the lines using the Heisenberg uncertainty principle ($\Delta \nu \approx 1/(2\pi \Delta t)$) one needs to estimate the lifetimes $\Delta t$ of the metastable excited states [57]. This is complicated by the fact that there are two channels for the decay of these states. These issues are currently being investigated. Such a broadening is consistent with the diffuseness of DIBs that is well-documented in the literature [1,6].

In sum, the clusters are predicted to be stable in their ground states at the low temperatures and densities endemic to the interstellar medium, and exhibit spectral lines consistent with certain DIBs. Subsequent research on such clusters is warranted.

**SUPPLEMENTARY INFORMATION**
A file listing the Cartesian coordinates of the optimized geometries all the clusters examined in this work and a display of their respective calculated IR spectra.

**ACKNOWLEDGEMENTS**
A preliminary version of this work has been presented as a poster entitled "Anionic Hydrogen Clusters ($H_{2n+1}^-$) as a Chemical Source of Diffuse Interstellar Bands and Possible Candidates for Hydrogen Storage" by Huang, L.; Matta, C. F.; Massa, L. at the *Sagamore XVII Conference on Charge, Spin and Momentum Densities (CSMD), International Union of Crystallography*; Kitayuzawa, Hokkai-do, Japan (15-20 July 2012) and at the *First Changsha International Workshop on Theoretical and Computational Chemistry with Materials*, Hunan Normal University, Changsha, China (8, 9 June 2012). The authors thank Professor Douglas Whitaker (*Mount Saint Vincent University*) for discussions. L. H. thanks the *Office of Naval Research*, for financial support. L. M.'s studies were funded by grants from the *U.S. Naval Research Laboratory* and PSC/CUNY. C. F. M. acknowledges the *Natural Sciences and Engineering Research Council of Canada* (NSERC), *Canada Foundation for Innovation* (CFI), and *Mount Saint Vincent University* for funding.

**REFERENCES**

[1] McCall BJ, Griffin RE. On the discovery of the diffuse interstellar bands. *Proc. Royal. Soc. A* 469, 2012.0604 (2013).

[2] Hartmann J. Investigations on the spectrum and orbit of delta Orionis. *Astrophys. J.* 19, 268-286 (1904).

[3] Hobbs LM, York DG, Thorburn JA, Snow TP, Bishof M, Friedman SD, McCall BJ, Oka T, Rachford B, Sonnentrucker P, Welty DE. Studies of the Diffuse Interstellar Bands. III. HD 183143. *Astrophys. J.* 705, 32-45 (2009).

[4] Ehrenfreund P, Cami Jiménez-Vicente J, Foing BH, Kaper L, van der Meer A, Cox N, d'Hendecourt L, Maier JP, Salama F, Sarre PJ, Snow TP, Sonnentrucker P. Detection of diffuse interstellar bands in the Magellanic Clouds. *Astrophys. J.* 576, L117-L120 (2002).

[5] Junkkarinen VT, Cohen RD, Beaver EA, Burbidge EM, Lyons RW, Madejski G. Dust and diffuse interstellar bands in the $z_a$ = 0.524 absorption system toward AO 0235+164. *Astrophys. J.* 614, 658-670 (2004).

[6] Sarre PJ. The diffuse interstellar bands: A major problem in astronomical spectroscopy. *J. Mol. Spectr.* 238, 1-10 (2006).

[7] Tielens AGGM. The molecular universe. *Rev. Mod. Phys.* 85, 1021-1081 (2013).

[8] *ESO Diffuse Interstellar Bands Large Exploration Survey (EDIBLES) – Merging Observations and Laboratory Data - March 29, 2016 (https://ntrs.nasa.gov/search.jsp?R=20160004200)*

[9] Campbell EK, Holz M, Gerlich D, Maier

JP. Laboratory confirmation of $C_{60}^+$ as the carrier of two diffuse interstellar bands. *Nature* 523, 322-323 (2015).

[10] Campbell EK, Holz M, Maier JP, Gerlich D, Walker GAH, Bohlender D. Gas phase absorption spectroscopy of $C_{60}^+$ and $C_{70}^+$ in a cryogenic ion trap: Comparison with astronomical measurements. *Astrophys J.* 822, 17 (7pp) (2016).

[11] Carruthers GR. Atomic and molecular hydrogen in interstellar space. *Space Sci. Rev.* 10, 459-482 (1970).

[12] Hill RN. Proof that the $H^-$ ion has only one bound state. *Phys. Rev. Lett.* 38, 643-646 (1977).

[13] Harrison SW, Massa LJ, Solomon P. Ion-induced dipole clusters. $Be_2^+He_2$. *Chem. Phys. Lett.* 16, 57-59 (1972).

[14] Harrison SW, Massa LJ, Solomon P. Binding energy and geometry of the hydrogen clusters $H_n^+$. *Nature* 245, 31-32 (1973).

[15] Harrison SW, Henderson GA, Massa LJ, Solomon P. Hartree-Fock bound states for molecule-ions $HeC_2^+$ and $HeC^+$. *Astrophys. J.* 189, 605-607 (1974).

[16] Sapse AM, Rayez-Meaume MT, Rayez JC, Massa LJ. Ion-induced dipole $H_n^-$ clusters. *Nature* 278, 332-333 (1979).

[17] Rayez JC, Rayez-Meaume MT, Massa LJ. Theoretical study of the hydride ($H_3^-$) cluster. *J. Chem. Phys.* 75, 5393-5397 (1981).

[18] Martinez OJr, Yang Z, Demarais NJ, Snow TP, Bierbaum VM. Gas-phase reactions of hydride anion, $H^-$. *Astrophys J.* 720, 173-177 (2010).

[19] McGuire BA. 2018 Census of interstellar, circumstellar, extragalactic, protoplanetary disk, and exoplanetary molecules. *Astrophys J. Supp.* 239, 17 (48 pp) (2018).

[20] Millar TJ, Walsh C, Field TA. Negative ions in space. *Chem. Rev.* 117, 1765-1795 (2017).

[21] Sarre PJ. The diffuse interstellar bands: a dipole-bound state hypothesis. *Mon. Not. R. Astron. Soc.* 313, L14-L16 (2000).

[22] Brillouin L. *Les Champs 'self-consistents` de Hartree et de Fock*. Hermann et Cie, Paris, (1934).

[23] Foresman JB, Head-Gordon M, Pople JA, Frisch MJ. Toward a systematic molecular orbital theory for excited state. *J. Phys. Chem.* 96, 135-149 (1992).

[24] Szabo A, Ostlund NS. *Modern Quantum Chemistry: Introduction to Advanced Electronic Structure Theory*. Dover Publications, Inc., New York, (1989).

[25] Bader RFW. *Atoms in Molecules: A Quantum Theory*. Oxford University Press, Oxford, U.K., (1990).

[26] Popelier PLA. *Atoms in Molecules: An Introduction*. Prentice Hall, London, (2000).

[27] Matta CF, Boyd RJ. *The Quantum Theory of Atoms in Molecules: From Solid State to DNA and Drug Design*. Wiley-VCH, Weinheim, (2007).

[28] Keith, T. A.; AIMAll – aim.tkgristmill.com: 2019.

[29] Matta CF, Huang L, Massa L. Characterization of a trihydrogen bond on the basis of the topology of the electron density. *J. Phys. Chem. A* 115, 12445-12450 (2011).

[30] Barbatti M, Nascimento MAC. On the formation mechanisms of hydrogen ionic clusters. *Braz. J. Phys.* 33, 792-797 (2003).

[31] Mathur D, Hasted JB. Odd and even numbered hydrogen ion clusters. *Nature* 280, 573-574 (1979).

[32] Clampitt R, Gowland L. Clustering of cold hydrogen gas on protons. *Nature* 223, 815-816 (1969).

[33] Calvo F, Yurtsever E. The quantum structure of anionic hydrogen clusters. *J. Chem. Phys.* 148, 102305 (pp. 1-8) (2018).

[34] Renzler M, Kuhn M, Mauracher A, Lindinger A, Scheier P, Ellis AM. Anionic hydrogen cluster ions as a new form of condensed hydrogen. *Phys. Rev. Lett.* 117, 273001 (pp. 1-4) (2016).

[35] Robinson K. *Spectroscopy: The Key to the Stars. Reading the Lines in Stellar Spectra*. Springer-Verlag London Ltd., London, (2007).

[36] Cox AN. *Allen's Astrophysical Quantities (Fourth Edition)*. Springer Science+Business Media, New York, (2002).

[37] Chaabouni H, Diana S, Nguyen T, Dulieu F. Thermal desorption of formamide and methylamine from graphite and amorphous water ice surfaces. *Astron. Astrophys.* 612, A47 (pp. 1-12) (2018).

[38] Turro NJ. *Modern Molecular Photochemistry*. University Science Books, Sausalito, CA, (1991).

[39] Wang W, Belyaev AK, Xu Y, Zhu A, Xiao C, Yang X-F. Observations of H3- and D3- from dielectric barrier discharge

plasmas. *Chem. Phys. Lett.* 377, 512-518 (2003).

[40] Gillespie RJ, Hargittai I. *The VSEPR Model of Molecular Geometry*. Allyn and Bacon, Boston, (1991).

[41] Bakhmutov VI. *Dihydrogen Bonds: Principles, Experiments, and Applications*. Wiley-Interscience, New Jersey, (2008).

[42] Parr RG, Yang W. *Density-Functional Theory of Atoms and Molecules*. Oxford University Press, Oxford, (1989).

[43] Scholl DS, Steckel JA. *Density Functional Theory: A Practical Introduction*. Wiley, Hoboken, New Jersey, (2009).

[44] Koch W, Holthausen MC. *A Chemist's Guide to Density Functional Theory, (Second Edition)*. Wiley-VCH, New York, (2001).

[45] Huang L, Matta CF, Massa L. Ion induced dipole clusters (3 ≤ n-odd ≤ 13): Density functional theory calculations of structure and energy. *J. Phys. Chem. A* 115, 12451-12458 (2011).

[46] Wolff M, Meyer J, Feldmann C. [$C_4MPyr$]$_2$[$Br_{20}$]: Ionic-liquid-based synthesis of a three-dimensional polybromide network. *Angew. Chem. Int. Ed.* 50, 4970-4973 (2011).

[47] Pichierri F. Structure and bonding in polybromide anions $Br^-(Br_2)_n$ (n = 1-6). *Chem. Phys. Lett.* 515, 116-121 (2011).

[48] Pichierri F. Binding of molecular hydrogen to halide anions: A computational exploration of eco-friendly materials for hydrogen storage. *Chem. Phys. Lett.* 519-520, 83-88 (2012).

[49] Herbig GH. Diffuse interstellar bands. 4 Region 4400-6850 Å. *Astrophys. J.* 196, 129-160 (1975).

[50] Jenniskens P, Désert F-X. A survey of diffuse interstellar bands (3800-8680 Å). *Astron. Astrophys. Suppl. Ser.* 106, 39-78 (1994).

[51] Halasinski TM, Salama F, Allamandola LJ. Investigation of the ultraviolet, visible, and near-infrared absorption spectra of hydrogenated polycyclic aromatic hydrocarbons and their cations. *Astrophys. J.* 628, 555-566 (2005).

[52] Tuairisg SÓ, Cami J, Foing BH, Sonnentrucker P, Ehrenfreund P. A deep echelle survey and new analysis of diffuse interstellar bands. *Astron. Astrophys. Suppl. Ser.* 142, 225-238 (2000).

[53] McCall BJ, Drosback MM, Thorburn JA, York DG, Friedman SD, Hobbs LM, Rachford BL, Snow TP, Sonnentrucker P, Welty DE. Studies of Diffuse Interstellar Bands. IV. The Nearly Perfect Correlation Between λλ6196.0 and 6613.6. *Astrophys. J.* 708, 1628-1638 (2010).

[54] Snow TP, Destree JD. The Diffuse Interstellar Bands in History and in the UV. *EAS (European Astronomical Society) Publications Series Vol. 46: PAHs and the Universe: A Symposium to Celebrate the 25th Anniversary of the PAH Hypothesis (C. Joblin and A.G.G.M. Tielens (eds))* 341-347 (2011).

[55] Kwok S. *Physics and Chemistry of the Interstellar Medium*. University Science Books, Sausalito, California, (2007).

[56] Rau ARP. The negative ion of hydrogen. *J. Astrophys. Astr.* 17, 113-145 (1996).

[57] Thomsen VBE. Why do spectral lines have a linewidth? *J. Chem. Edu.* 72, 616-618 (1995).

**Table 1**
Energies corresponding to optimized geometries of the anion induced dipolar hydrogen clusters, their highest vibrational frequencies, and predicted lifetimes.

|  | Point Group | $E_{MP2}$ (au) | HOMO (au) | LUMO (au) | HOMO-LUMO Gap (au) | Binding energy[a] (eV) | Highest vibrational freq. (Hz) | Lifetime ($\tau$) (s) | |
|---|---|---|---|---|---|---|---|---|---|
|  |  |  |  |  |  |  |  | Room temp. (298 K)[c] | Interstellar cloud temp. (20 K)[c] |
| $H_3^-$ | $C_v$ | -1.6825 | -0.0463 | 0.1465 | -0.1928 | -0.05 | $1.40 \times 10^{14}$ | $5.2 \times 10^{-14}$ | $4.5 \times 10^{-2}$ |
| $H_5^-$ | $C_s$ | -2.8498 | -0.0480 | 0.1376 | -0.1857 | -0.12 | $1.39 \times 10^{14}$ | $7.5 \times 10^{-13}$ | $9.0 \times 10^{15}$ |
| $H_7^-$ | $C_{3v}$ | -4.0155 | -0.0486 | 0.1287 | -0.1774 | -0.15 | $1.38 \times 10^{14}$ | $2.1 \times 10^{-12}$ | $4.9 \times 10^{22}$ |
| $H_9^-$ | $C_1$ | -5.1756 | -0.0428 | 0.1277 | -0.1706 | -0.02 | $1.38 \times 10^{14}$ | $1.5 \times 10^{-14}$ | $3.4 \times 10^{-10}$ |
| $H_{11}^-$ | $C_{4v}$ | -6.3460 | -0.0484 | 0.1135 | -0.1619 | -0.17 | $1.38 \times 10^{14}$ | $6.1 \times 10^{-12}$ | $2.5 \times 10^{29}$ |
| $H_{13}^-$ | $C_1$ | -7.5115 | -0.0481 | 0.1145 | -0.1626 | -0.19 | $1.39 \times 10^{14}$ | $1.3 \times 10^{-11}$ | $2.3 \times 10^{34}$ |
| $H_{15}^-$ | $C_1$ | -8.6774 | -0.0486 | 0.1130 | -0.1615 | -0.22 | $1.39 \times 10^{14}$ | $4.5 \times 10^{-11}$ | $2.8 \times 10^{42}$ |

(a) See Eq. (1) for definition.
(b) Thermal energy at room temperature is ~ 0.0257 eV and at the interstellar dust clouds temperature (20 K) is ~ 0.0017 eV.

**Table 2**
QTAIM atomic charges in atomic units (au). H1 is the central anion in all clusters.

| Atom | Cluster | | | | | | |
|---|---|---|---|---|---|---|---|
| | $H_3^-$ | $H_5^-$ | $H_7^-$ | $H_9^-$ | $H_{11}^-$ | $H_{13}^-$ | $H_{15}^-$ |
| H1 | -0.953 | -0.942 | -0.937 | -0.936 | -0.936 | -0.918 | -0.909 |
| H2 | +0.105 | +0.083 | +0.067 | +0.059 | +0.040 | +0.034 | +0.031 |
| H3 | -0.148 | -0.110 | -0.087 | -0.080 | -0.042 | -0.036 | -0.032 |
| H4 | | +0.083 | +0.067 | +0.026 | +0.051 | +0.053 | +0.033 |
| H5 | | -0.110 | -0.087 | -0.038 | -0.066 | -0.069 | -0.047 |
| H6 | | | +0.067 | +0.017 | +0.051 | +0.033 | +0.037 |
| H7 | | | -0.087 | -0.028 | -0.066 | -0.044 | -0.049 |
| H8 | | | | +0.023 | +0.051 | +0.051 | +0.028 |
| H9 | | | | -0.042 | -0.066 | -0.066 | -0.037 |
| H10 | | | | | +0.051 | +0.040 | +0.055 |
| H11 | | | | | -0.066 | -0.054 | -0.070 |
| H12 | | | | | | +0.077 | +0.078 |
| H13 | | | | | | -0.100 | -0.100 |
| H14 | | | | | | | +0.062 |
| H15 | | | | | | | -0.080 |

**Table 3**
Comparison of the calculated lines (CIS-MP2/aug-cc-pVTZ//CIS/aug-cc-pVTZ) and observed DIBs.[†]

| Band number | Cluster | Oscillator Strength ($f$) | $\lambda_{max}$(CIS+MP2) (nm) | $\lambda_{max}$(DIBS)[(a)] (nm) | $\Delta\lambda_{max}$[(b)] (nm) |
|---|---|---|---|---|---|
| 1 | $H_9^-$ | 0.0435 | 440.41 | 442.89 (0.135) | -2.48 |
| **2*** | $H_{13}^-$ | **0.0043** | **443.18** | **442.89 (0.135)** | **0.29** |
| 3 | $H_{15}^-$ | 0.0070 | 445.20 | 442.89 (0.135) | 2.31 |
| **4*** | $H_{13}^-$ | **0.0023** | **449.07** | **450.18 (0.070)** | **-1.11** |
| 5 | $H_{11}^-$ | 0.0000 | 451.54 | 450.18 (0.070) | 1.36 |
| **6*** | $H_9^-$ | **0.0073** | **457.46** | **459.50 (0.300)**[(c)] | **-2.04** |
| 7 | $H_{11}^-$ | 0.0071 | 474.91 | 472.71 (0.071) | 2.20 |
| 8 | $H_9^-$ | 0.0441 | 474.91 | 476.17 (0.058) | -1.26 |
| 9 | $H_7^-$ | 0.0004 | 477.85 | 478.01 (0.029) | -0.16 |
| 10 | $H_7^-$ | 0.0003 | 477.93 | 478.01 (0.029) | -0.08 |
| 11 | $H_5^-$ | 0.0000 | 479.65 | 478.01 (0.029) | 1.64 |
| **12*** | $H_{13}^-$ | **0.0130** | **482.69** | **482.40 (0.20)** | **0.29** |
| 13 | $H_9^-$ | 0.0348 | 488.14 | 488.18 (0.065) | -0.04 |
| **14*** | $H_{15}^-$ | **0.0028** | **493.54** | **496.40 (0.004)** | **-2.86** |
| **15*** | $H_{11}^-$ | **0.0018** | **505.23** | **503.91 (0.113)**[(c)] | **1.32** |
| 16 | $H_{11}^-$ | 0.0018 | 505.24 | 503.91 (0.113)[(c)] | 1.33 |
| 17 | $H_{15}^-$ | 0.0033 | 508.57 | 510.97 (0.240) | -2.40 |
| **18*** | $H_{13}^-$ | **0.0030** | **510.97** | **510.97 (0.240)** | **0.00** |
| 19 | $H_3^-$ | 0.0134 | 530.75 | 536.21 (0.090) | -5.46 |
| 20 | $H_5^-$ | 0.0226 | 549.60 | 549.41 (0.014) | 0.19 |
| 21 | $H_7^-$ | 0.0256 | 584.24 | 584.42 (0.022) | -0.18 |
| 22 | $H_{15}^-$ | 0.0139 | 598.36 | 598.20 (0.032)[(c)] | 0.16 |
| 23 | $H_{13}^-$ | 0.0167 | 606.70 | 606.04 (0.067)[(d)] | 0.66 |
| 24 | $H_{11}^-$ | 0.0213 | 612.92 | 613.98 (0.003) | -1.06 |
| 25 | $H_9^-$ | 0.0061 | 619.77 | 619.62 (0.033) | 0.15 |

[†] The entries in this table are sorted as a function of increasing wavelength.
(a) Experimental DIBS data taken from Ref. [50]. Values in brackets convey the observational uncertainties. (b) $\Delta\lambda_{max} \equiv \lambda_{max}$(CIS+MP2) $- \lambda_{max}$(DIB). (c) Uncertain or questionable DIB. (d) Probable DIB.
* Definitive matches are labeled with an asterisk, and are likewise highlighted in bold.

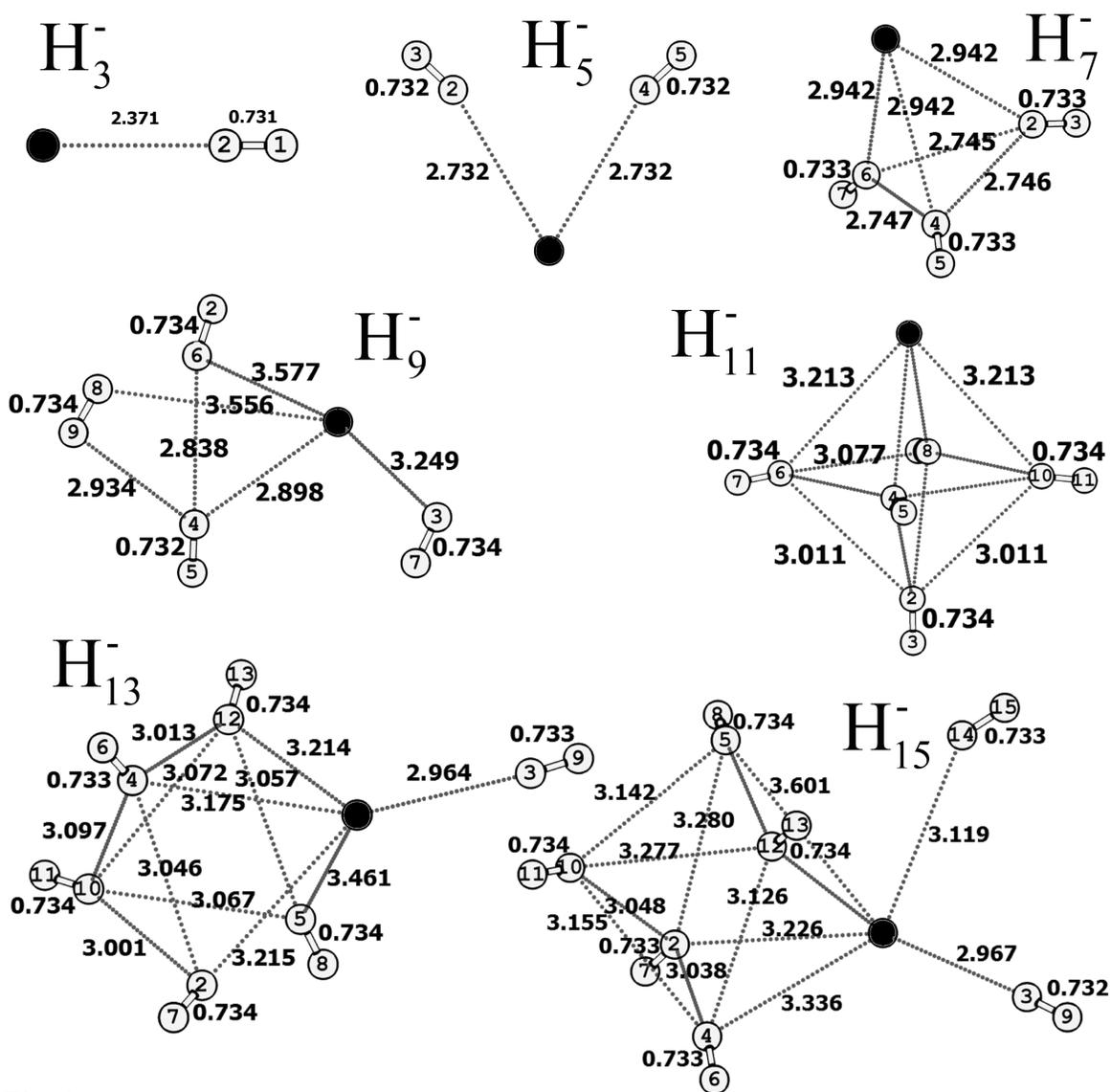

**Fig. 1**
Ball-and-stick representations of the optimized geometries of all the clusters with the atom numbering scheme and the optimized bond lengths (in Å). Covalent H–H bonds are depicted as light gray cylinders, weak interactions (either trihydrogen bonds or weak H–H bonds) are depicted as dotted lines. In every cluster, the black sphere represents the hydride anion (H⁻).

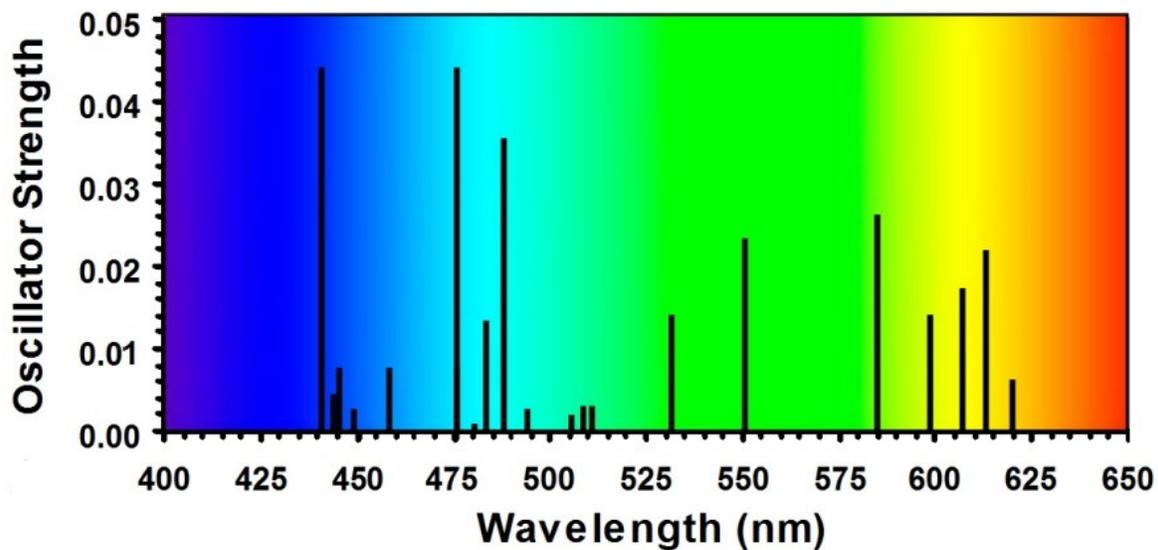

**Fig. 2**
Calculated visible spectra associated with certain anionic hydrogen clusters.

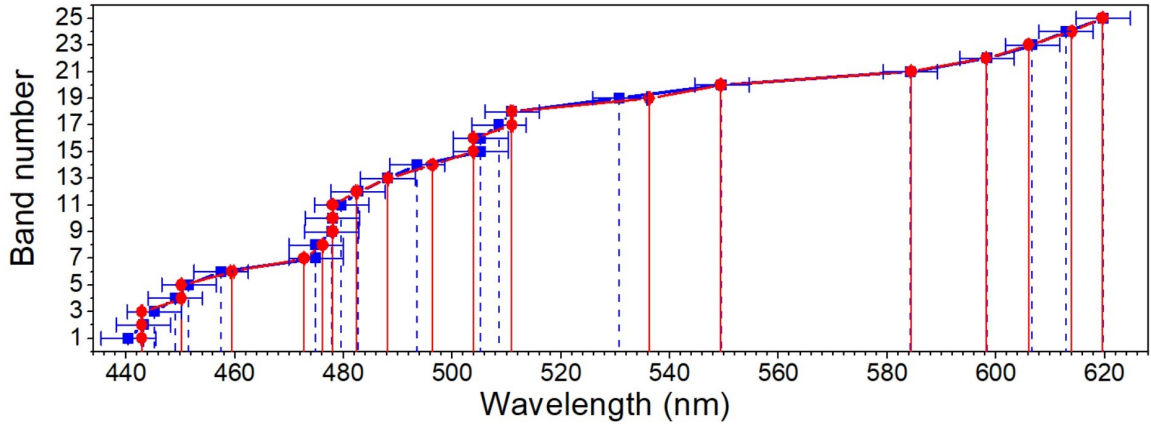

**Fig. 3**
Calculated wavelengths (red), along with estimated uncertainties (*ca.* ± 5 nm), compared with the observed DIB wavelengths (blue). Uncertainties linked to the latter are adopted from Ref. [50], and listed in Table 3. Band numbers correspond to their listing in Table 3.

# Supplementary Information for the Manuscript: Anionic Hydrogen Clusters as a Source of Diffuse Interstellar Bands (DIBs)

Lulu Huang,[1] Chérif F. Matta*,[2-5] Daniel Majaess,[2] Tina Harriott,[2,6] Joseph Capitani,[7] and Lou Massa*[1]

[1] *Hunter College and the Graduate School, City University of New York, New York, NY 10065, USA.* [2] *Department of Chemistry and Physics, Mount Saint Vincent University, Halifax, Nova Scotia, Canada B3M2J6.* [3] *Department of Chemistry, Saint Mary's University, Halifax, Nova Scotia, Canada B3H3C3.* [4] *Dép. de chimie, Université Laval, Québec, Québec, Canada G1V 0A6.* [5] *Department of Chemistry, Dalhousie University, Halifax, Nova Scotia, Canada B3H4J3.* [6] *Department of Mathematics and Statistics, Mount Saint Vincent University, Halifax, Nova Scotia, Canada B3M2J6.* [7] *Chemistry and Biochemistry Department, Manhattan College, Riverdale, NY 10471, USA.*

\* Correspondence: cherif.matta@msvu.ca, lmassa@hunter.cuny.edu

## Cartesian Coordinates and IR spectra of the Clusters

**$H_3^-$**
Charge = -1 Multiplicity = 1
H,0,0.,0.,-1.2774569187
H,0,0.,0.,-0.5469063546
H,0,0.,0.,1.8243632733

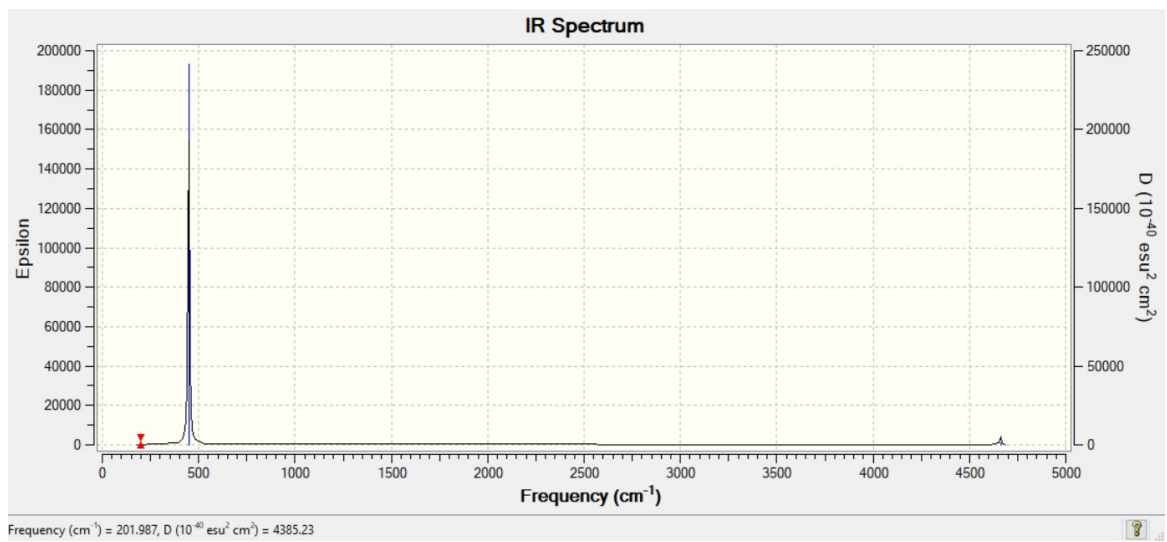

Frequency (cm$^{-1}$) = 201.987, D (10$^{-40}$ esu$^2$ cm$^2$) = 4385.23

**$H_5^-$**
Charge = -1 Multiplicity = 1

H,0,2.0709987417,-0.1794536735,0.
H,0,-0.3883399036,-1.3700473501,0.
H,0,-0.9345895161,-1.8573223519,0.
H,0,-0.1468913493,1.416423031,0.
H,0,-0.6011779727,1.9904003344,0.

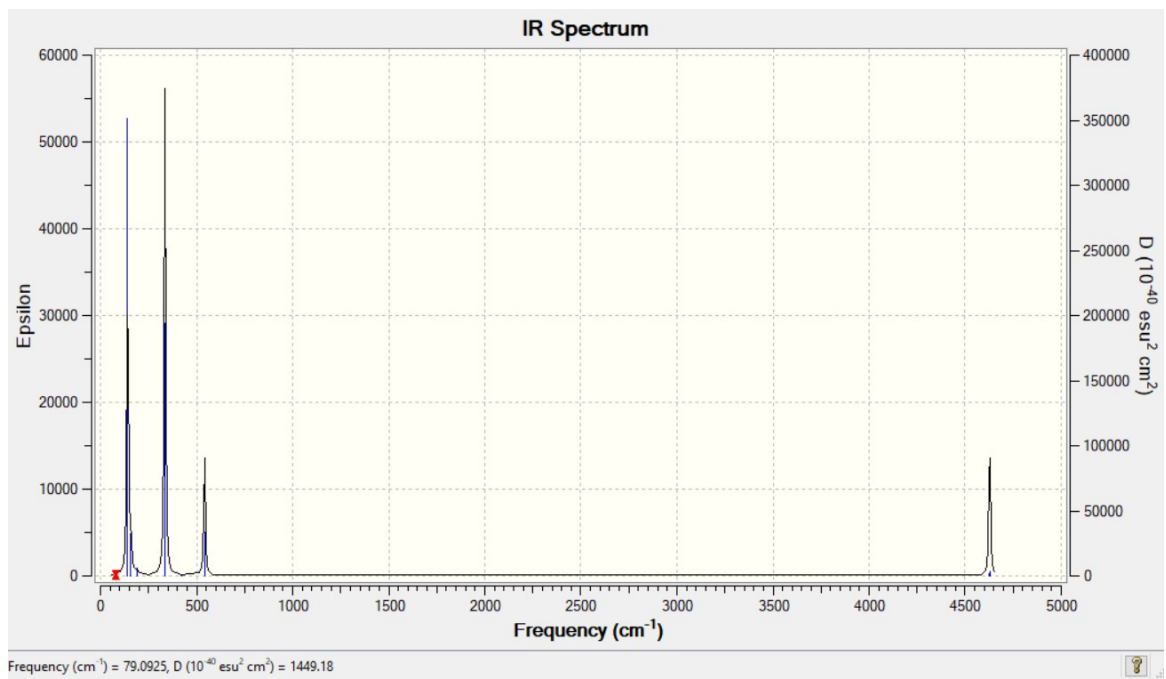

## H₇⁻

```
Charge = -1 Multiplicity = 1
 H,0,-0.0016356267,0.0016426527,2.2758008996
 H,0,-0.3962732166,1.534102341,-0.2038246506
 H,0,-0.5570193784,2.1560019135,-0.5567180094
 H,0,1.52866991,-0.4244205231,-0.2009721911
 H,0,2.1467660835,-0.595863967,-0.5555336147
 H,0,-1.1308649091,-1.1112715858,-0.2025008215
 H,0,-1.5896428627,-1.5601908312,-0.5562516123
```

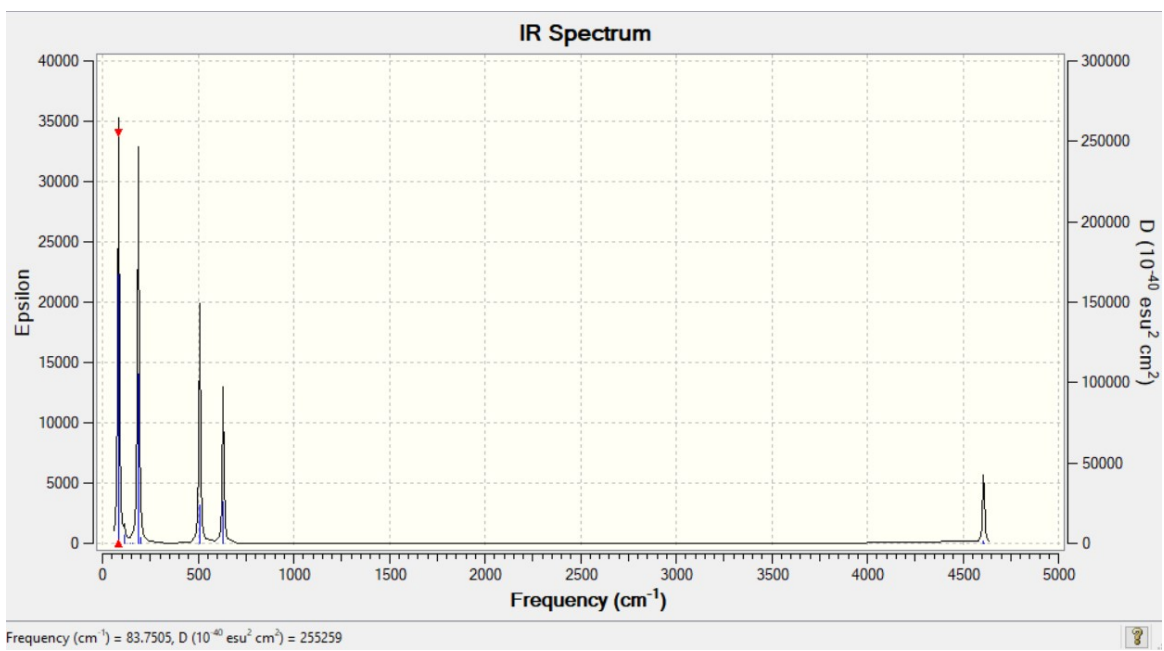

**H₉⁻**
```
Charge = -1 Multiplicity = 1
 H,0,0.8546679214,-1.100259905,-0.6213603689
 H,0,-2.9235637709,-1.5538753088,-0.8432226843
 H,0,4.0557938211,-0.5614589322,-0.481602327
 H,0,-0.80727877,0.1143634303,1.4191463341
 H,0,-0.7201714228,0.1534564855,2.1450649281
 H,0,-2.7035608163,-1.1642554075,-0.2611017416
 H,0,4.1526549755,-0.1594816581,0.1252731828
 H,0,-1.002941,1.9055455191,-1.0227680696
 H,0,-0.9056009379,2.3659667768,-0.4594292534
```

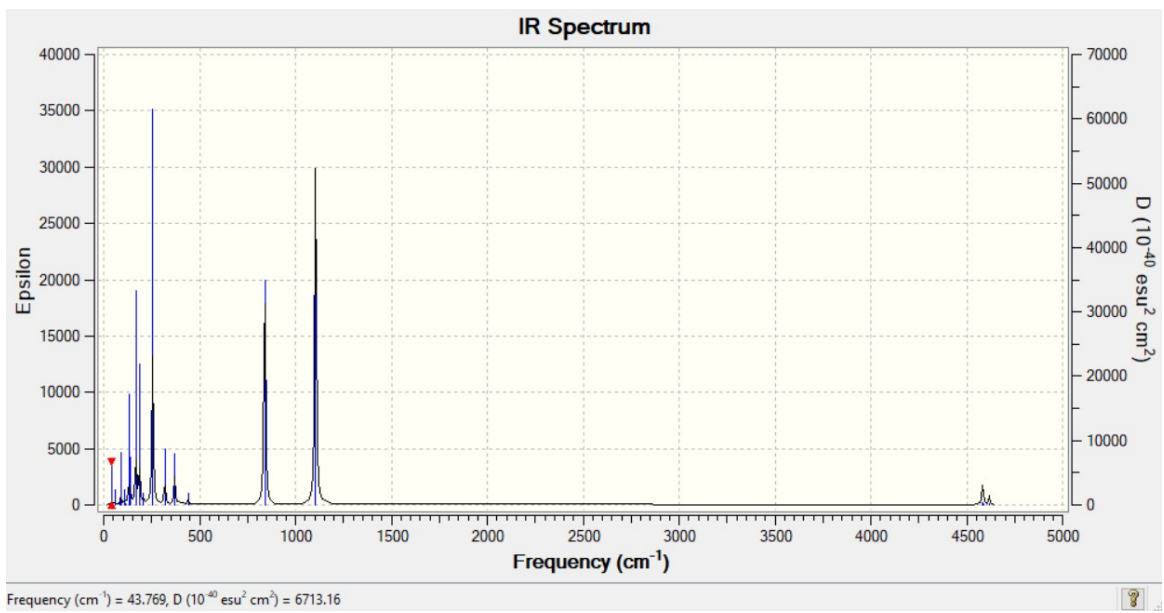

# $H_{11}^-$
```
Charge = -1 Multiplicity = 1
 H,0,0.0002331591,-0.0000321656,-2.635163396
 H,0,0.0000086568,-0.0002595767,1.8099685362
 H,0,0.0000554593,-0.0001357407,2.5438643569
 H,0,-1.5443758481,-1.5327267382,-0.2713529704
 H,0,-2.2059010087,-1.8290808443,-0.1585586195
 H,0,1.5329836263,-1.5442178059,-0.2712372389
 H,0,1.8291651856,-2.2058244999,-0.1584357001
 H,0,1.5442325922,1.5330926559,-0.2711256062
 H,0,2.2059055199,1.8291808403,-0.158437372
 H,0,-1.5331802392,1.5441405608,-0.2711266997
 H,0,-1.8291281032,2.2058653144,-0.1583952903
```

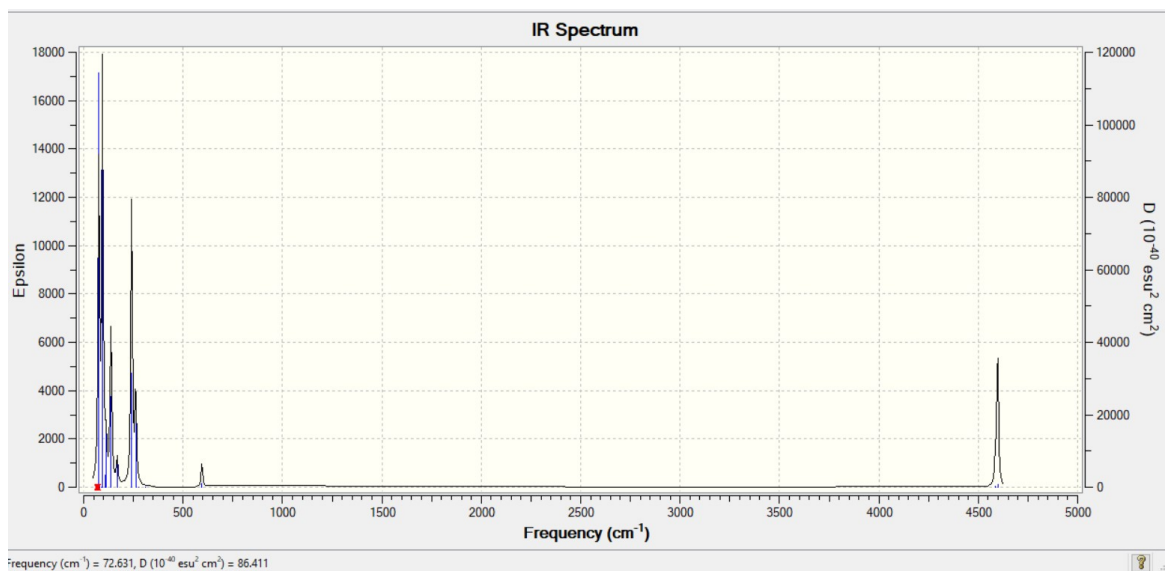

# $H_{13}^-$
```
Charge = -1 Multiplicity = 1
 H,0,-1.223431,-0.537884,1.583488
 H,0,1.36908,1.362942,1.558661
 H,0,-3.935059,-0.145429,0.452178
 H,0,1.703655,-1.609277,0.981691
 H,0,-0.62095,1.653172,-1.027503
 H,0,2.058384,-2.232473,1.135246
 H,0,1.920602,1.575343,1.993114
 H,0,-0.807508,2.361061,-0.975509
 H,0,-4.660754,-0.155989,0.349819
 H,0,2.228839,0.531295,-1.193984
 H,0,2.718462,0.355969,-1.711895
 H,0,-0.190014,-1.356808,-1.347756
 H,0,-0.561308,-1.801922,-1.797551
```

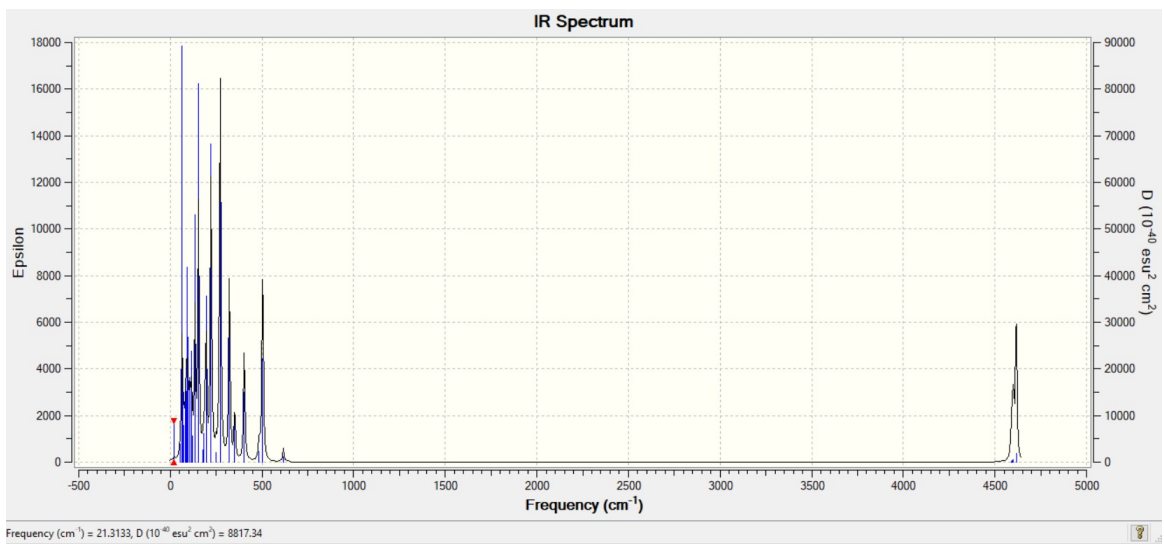

## H<sub>15</sub><sup>-</sup>

```
Charge = -1 Multiplicity = 1
 H,0,1.176362,-0.289762,-1.366944
 H,0,-1.992403,-0.102359,-1.943492
 H,0,3.501935,-1.779047,-0.282205
 H,0,-1.133715,-2.258881,0.016129
 H,0,-0.919624,2.274254,0.046697
 H,0,-0.995403,-2.969891,0.13157
 H,0,-2.510068,-0.323663,-2.413355
 H,0,-1.119377,2.831138,-0.387675
 H,0,4.118224,-2.158686,-0.17177
 H,0,-3.018128,0.104665,0.918987
 H,0,-3.474655,-0.159721,1.429261
 H,0,0.115577,-0.067267,1.862912
 H,0,0.576627,0.027681,2.42566
 H,0,2.524108,2.246501,-0.151885
 H,0,3.15054,2.625038,-0.113891
```

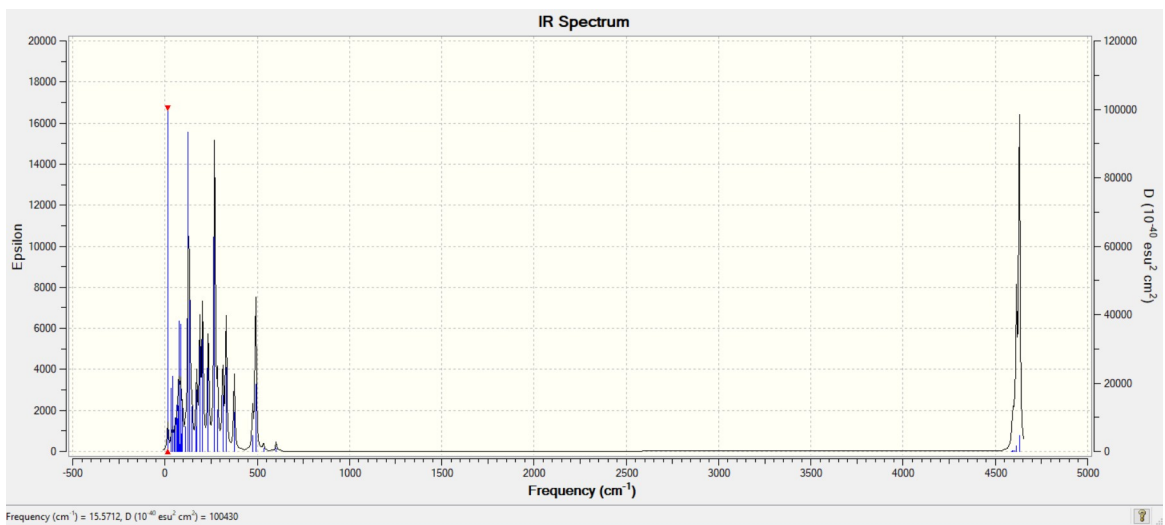

# Anionic Hydrogen Clusters ($H_{2n+1}^-$) as a Chemical Source of Diffuse Interstellar Bands and Possible Candidates for Hydrogen Storage

Lulu Huang,[(1)] Chérif F. Matta,[(2)] and Lou Massa[(3)]

[(1)] Center for Computational Materials Science, Naval Research Laboratory, Washington, DC 20375-5341, USA.
[(2)] Department of Chemistry, Mount Saint Vincent University, Halifax, Nova Scotia, Canada B3M 2J6.
[(3)] Hunter College and the Graduate School, City University of New York, New York, NY 10065, USA.

© Huang, Massa, Matta (2012)

## SUMMARY

One of the oldest outstanding mysteries of astronomical spectroscopy is the nature of the chemical source of the **"diffuse interstellar bands" (DIB)**.[1,2] The spectra of stars observed through interstellar clouds exhibit these characteristic absorption bands. The chemical source of DIB remains an unsettled question despite many plausible proposals. It is suggested that a possible chemical source of DIB is a family of ion-induced dipole clusters of the general formula consisting of a central anion ($H^-$) holding $nH_2$ ligands ($1 \leq n \leq 7$).[3-10] Configuration interaction calculations (CI) demonstrate **(a)** that these clusters are stable, and **(b)** that their spectra exhibit 25 lines in the visible region all coinciding, within carefully estimated uncertainty bars, with the observed DIB. Furthermore, these clusters may also provide a basis for an efficient type of hydride-centered hydrogen storage material in view of their close similarity to recently reported halide-centered hydrogen clusters.[11,12] The clusters are primarily held by a trihydrogen bonding interaction between the $H_2$ ligands and the central anion.[9] In addition, it is shown that weak van der Waals' type interactions, manifested in well-characterized H...H bond paths between neighbouring $H_2$ ligands, contribute to the stability of these clusters.[8-10] These closed-shell interactions only exist in the densities obtained from calculations that properly account for dispersion such as CI.

## COMPUTATIONS

The triple-zeta quality aug-cc-pVTZ basis set has been used in all calculations reported in this paper, and all geometries were completely optimized through an unconstrained energy minimization at the CIS/aug-cc-pVTZ. The level of theory used in all reported wavelength ($\lambda_{max}$) calculations is CIS-MP2/aug-cc-pVTZ//CIS/aug-cc-pVTZ, a level that has been found to reproduce UV-Vis spectra accurately.

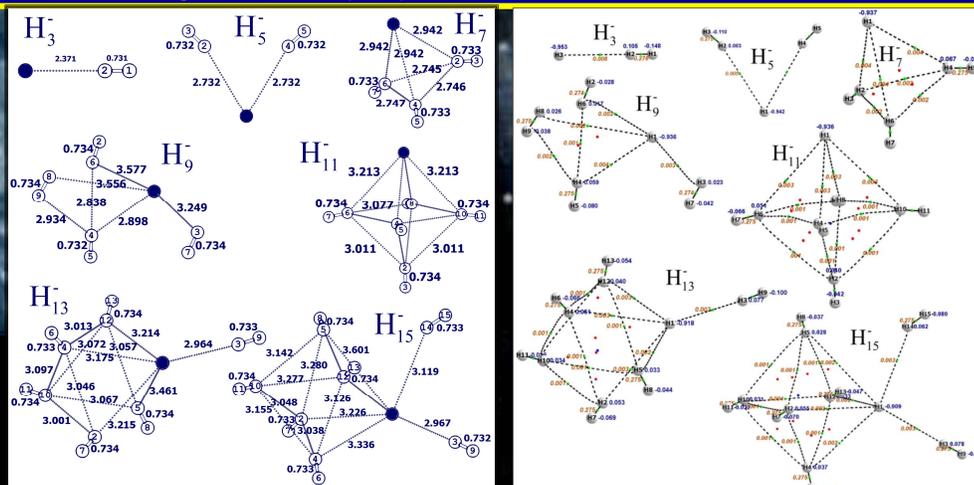

Ball-and-stick representation of the optimized geometries of all the clusters reported in this paper along with the atom numbering scheme and the optimized bond lengths (in Å). Covalent H–H bonds are depicted as light gray cylinders, weak interactions (either trihydrogen bonds or weak H–H bonds) are depicted as dotted lines. In every cluster, the black sphere represents the hydride anion ($H^-$).

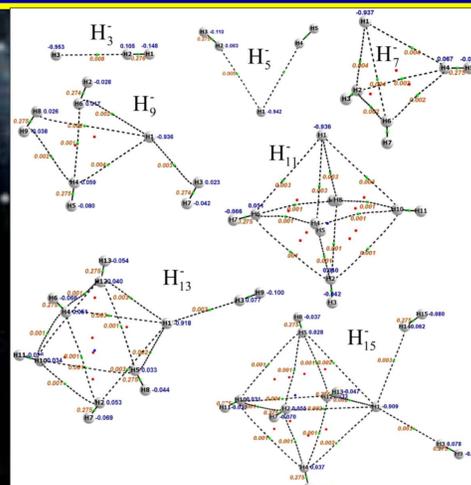

Bond paths defining the molecular graphs of the studied clusters, along with atomic charges and the electron density at the bond critical points (au). Some symmetry non-unique atomic and bond properties are omitted for clarity. Green spheres indicate bond critical points, red spheres for ring critical points, and blue spheres for cage critical points.

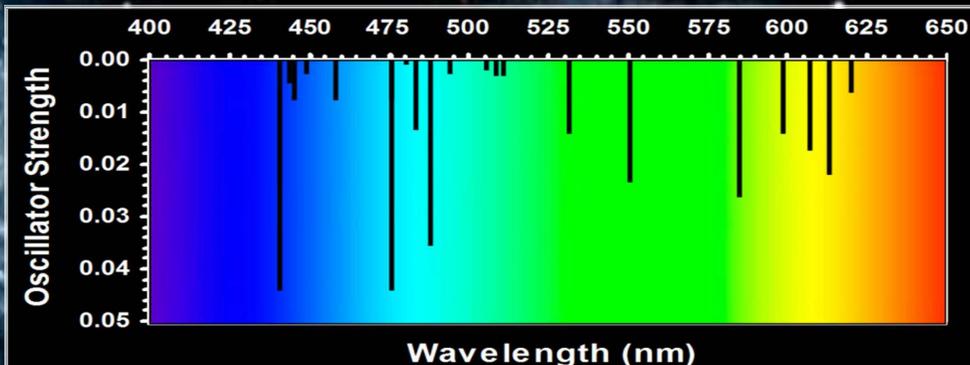

**TOP:** Calculated spectra that match diffuse interstellar bands (DIB).

**RIGHT:** Calculated wavelengths (blue), along with the associated uncertainty bars of the calculations (ca. ± 5 nm), compared with the experimental wavelengths (red). The uncertainties of the experimental DIB are taken from Ref. 2 and are listed in the **Table**. Band numbers correspond to their listing in the **Table**.

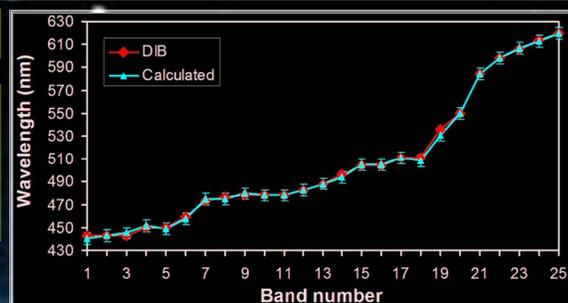

| Band number | Cluster | Oscillator Strength (f) | $\lambda_{max}$(CIS+MP2) (nm) | $\lambda_{max}$(DIB)[(a)] (nm) | $\Delta\lambda_{max}$[(b)] (nm) |
|---|---|---|---|---|---|
| 1 | $H_3^-$ | 0.0435 | 440.41 | 442.89 (0.135) | -2.48 |
| 2 | $H_{13}^-$ | 0.0043 | 443.18 | 442.89 (0.135) | 0.29 |
| 3 | $H_{15}^-$ | 0.0070 | 445.20 | 442.89 (0.135) | 2.31 |
| 4 | $H_{13}^-$ | 0.0023 | 449.07 | 450.18 (0.070) | -1.11 |
| 5 | $H_{11}^-$ | 0.0000 | 451.54 | 450.18 (0.070) | 1.36 |
| 6 | $H_9^-$ | 0.0073 | 457.46 | 459.50 (0.300)[(c)] | -2.04 |
| 7 | $H_{11}^-$ | 0.0071 | 474.91 | 472.71 (0.071) | 2.20 |
| 8 | $H_9^-$ | 0.0441 | 474.91 | 476.17 (0.058) | -1.26 |
| 9 | $H_7^-$ | 0.0004 | 477.85 | 478.01 (0.029) | -0.16 |
| 10 | $H_7^-$ | 0.0003 | 477.93 | 478.01 (0.029) | -0.08 |
| 11 | $H_5^-$ | 0.0000 | 479.65 | 478.01 (0.029) | 1.64 |
| 12 | $H_{13}^-$ | 0.0130 | 482.69 | 482.40 (0.20) | 0.29 |
| 13 | $H_9^-$ | 0.0348 | 488.14 | 488.18 (0.065) | -0.04 |
| 14 | $H_{15}^-$ | 0.0028 | 493.54 | 496.40 (0.004) | -2.86 |
| 15 | $H_{11}^-$ | 0.0018 | 505.23 | 503.91 (0.113)[(e)] | 1.32 |
| 16 | $H_{11}^-$ | 0.0018 | 505.24 | 503.91 (0.113)[(c)] | 1.33 |
| 17 | $H_{15}^-$ | 0.0033 | 508.57 | 510.97 (0.240) | -2.40 |
| 18 | $H_{13}^-$ | 0.0030 | 510.97 | 510.97 (0.240) | 0.00 |
| 19 | $H_3^-$ | 0.0134 | 530.75 | 536.21 (0.090) | -5.46 |
| 20 | $H_5^-$ | 0.0226 | 549.60 | 549.41 (0.014) | 0.19 |
| 21 | $H_7^-$ | 0.0256 | 584.24 | 584.42 (0.022) | -0.18 |
| 22 | $H_{15}^-$ | 0.0139 | 598.36 | 598.20 (0.032)[(c)] | 0.16 |
| 23 | $H_{13}^-$ | 0.0167 | 606.70 | 606.04 (0.067)[(d)] | 0.66 |
| 24 | $H_{11}^-$ | 0.0213 | 612.92 | 613.98 (0.003) | -1.06 |
| 25 | $H_9^-$ | 0.0061 | 619.77 | 619.62 (0.033) | 0.15 |

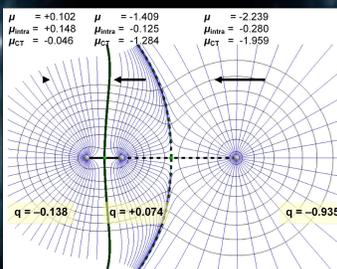

Representations of the electron density of $H_3^-$. **Upper panel:** Electron density contour lines 0.001, 0.002, 0.004, 0.008, 0.02, 0.04, 0.08 and 0.20 au with the associated gradient vector field lines, bond paths (covalent bond paths in solid lines and the weak trihydrogen bond path in dotted lines). Atomic contributions to the cluster's dipole moment are depicted by the arrows (proportional in length to the magnitude): $\mu$ is the total contribution of an atom = $\mu_{intra} + \mu_{charge\ transfer}$, "q" are the atomic charges in au.

## REFERENCES

[1] E. Wilson, *Chem. Eng. News* **88(49)**, 41 (2010).
[2] P. Jenniskens and F.-X. Désert, *Astron. Astrophys. Suppl. Ser.* **106**, 39 (1994).
[3] S. W. Harrison, L. J. Massa, and P. Solomon, *Chem. Phys. Lett.* **16**, 57 (1972).
[4] S. W. Harrison, L. J. Massa, and P. Solomon, *Nature* **245**, 31 (1973).
[5] S. W. Harrison, G. A. Henderson, L. J. Massa, and P. Solomon, *Astrophys. J.* **189**, 605 (1974).
[6] A. M. Sapse, M. T. Rayez-Meaume, J. C. Rayez, and L. J. Massa, *Nature* **278**, 332 (1979).
[7] J. C. Rayez, M. T. Rayez-Meaume, and L. J. Massa, *J. Chem. Phys.* **75**, 5393 (1981).
[8] L. Huang, C. F. Matta, and L. Massa, *J. Phys. Chem. A* **115**, 12451 (2011).
[9] C. F. Matta, L. Huang, and L. Massa, *J. Phys. Chem. A* **115**, 12445 (2011).
[10] L. Huang, C. F. Matta, and L. Massa, *submitted* (2012).
[11] F. Pichierri, *Chem. Phys. Lett.* **519-520**, 83 (2012).
[12] F. Pichierri, *Chem. Phys. Lett.* **515**, 116 (2011).

## ACKNOWLEDGEMENTS

The authors thank **Professor Christopher Cramer** (University of Minnesota) for his suggestion that the anionic hydrogen clusters may be of useful as hydrogen storage complexes. L. H. thanks the **Office of Naval Research**, for financial support. L. M.'s studies were funded by grants from the **U.S. Naval Research Laboratory** and **PSC/CUNY**. C. F. M. acknowledges the **Natural Sciences and Engineering Research Council of Canada (NSERC)**, **Canada Foundation for Innovation (CFI)**, and **Mount Saint Vincent University** for funding.